\documentclass[aps,showpacs,prl,twocolumn,floatfix]{revtex4}
\usepackage{amsmath,amssymb,graphicx,bm}
\begin{document}



\title{Where is the Shot Noise of a Quantum Point Contact?}

\author{
Frederick Green${}^{1,2}$,
Jagdish S. Thakur${}^3$,
and Mukunda P. Das${}^1$\\
${}^1$  {\em Department of Theoretical Physics,
Institute of Advanced Studies,
The Australian National University,
Canberra, ACT 0200, Australia.}\\
${}^2$ {\em School of Physics, The University of New South Wales,
Sydney, NSW 2052, Australia.}\\
${}^3$ {\em Department of Electrical and Computer Engineering,
Wayne State University, Detroit, Michigan 48202, USA.}
}

\bigskip
\begin{abstract}
Reznikov {\em et al.} (Phys. Rev. Lett. {\bf 75}, 3340 (1995))
have presented definitive observations of nonequilibrium noise
in a quantum point contact. Especially puzzling is the
``anomalous'' peak structure of the excess noise
measured at constant current; to date it remains unexplained.
We show that their experiment directly reveals the deep link
between conservation principles in the electron gas
and its low-dimensional, mesoscopic behavior.
Key to that connection are gauge invariance and the
compressibility sum rule. These are central
not only to the experiment of Reznikov {\em et al.}
but to the very nature of all mesoscopic transport.
\end{abstract}


\maketitle

Microscopic conservation laws are the {\em sine qua non}
of transport and fluctuation physics.
Their dominance is even more apparent in the passage to
sub-micrometer electronics.
In the following, we bring to light the direct governing role
of conservation in low-dimensional mesoscopic conduction,
with immediate experimental consequences. Indeed,
these have already been observed
\cite{rez}.

The primary global statements of conservation
are the electron-gas sum rules
\cite{pinoz}.
Their cardinal significance as conserving relations is that
they must be satisfied
{\em automatically}
by credible models of mesoscopic transport
\cite{thakur}.

This paper details
the particular and striking interplay of the compressibility sum rule
and dynamical electron relaxation. That interplay dominates the
form of the carrier fluctuations (noise) of a driven
quasi-one-dimensional quantum point contact,
so completely as to dictate the shape of its nonequilibrium
noise spectrum. Our results have
far-reaching implications for understanding transport
in mesoscopic electron devices.

Some years ago Reznikov, Heiblum, Shtrikman, and Mahalu
\cite{rez}
performed a landmark experiment on nonequilibrium noise
in a quantum point contact (QPC). The electron
density was freely adjusted via a gate voltage $V_g$,
sweeping the carriers from their low-density classical regime
up to high density and degeneracy.
At fixed values of source-drain voltage $V$ across the QPC,
a regular sequence of peaks appeared in the noise power
spectrum as the channel's carrier density was systematically
increased. Analogous features were also seen by Kumar {\em et al.}
\cite{kumar}.
The behavior at fixed $V$
is predicted by the noninteracting one-electron
picture of coherent ballistic conduction,
first applied to QPC shot noise by Khlus
\cite{khlus}.
It has since been refined and redefined by Landauer,
B\"uttiker, Imry, and many others;
for an authoritative review see Ref.
\onlinecite{blbu}
and citations in it.

A major innovation of Reznikov {\em et al.}
was to measure the nonequilibrium noise for fixed levels
of source-drain {\em current} $I$ through the QPC,
as well as their standard measurements at fixed
$V$.
A surprise ensued. Far from the anticipated
strictly monotonic falloff of spectral strength
\cite{blbu},
the data exhibited a series of pronounced noise maxima at the
lowest subband energy threshold.

\begin{figure}
  \centering
  \includegraphics[height=70mm]{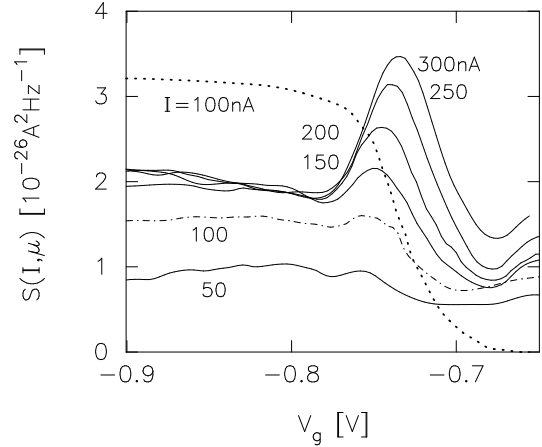}
  \caption{\label{f1}
Nonequilibrium current noise in a quantum point contact at 1.5K,
measured by Reznikov {\em et al}.
\cite{rez}
as a function of gate bias, at fixed source-drain current.
Dotted line: the most widely adopted theoretical noise model
\cite{blbu}
typically produces a strictly monotonic noise signal at
the first subband energy threshold.
That model totally fails to predict the very strong noise
peaks actually observed at threshold.}
\end{figure}

Figure \ref{f1} shows the constant-current data of Ref.
\onlinecite{rez}.
At higher source-drain currents, a robust peak
structure emerges in the noise as a function of gate bias,
precisely at the location of the first subband threshold.
At $I = 100$nA, we reproduce the
conventionally predicted shot-noise curve
\cite{rez,blbu}.

Reznikov {\em et al}. remark that
``the peaks near ${\cal T}_1 = 1/2$ [{\em ie} at threshold]
are not expected''
\cite{rez}.
This is an understatement, for the observed excess-noise maxima
contradict the accepted theoretical predictions,
based on shot noise. Something quite uncommon and
apparently inexplicable is going on.

In the nine years since publication of that
experiment, its astonishing outcome has not been revisited.
Yet the anomalous signatures are first-hand evidence
that a deep knowledge gap exists in our physical understanding
of mesoscopics. This has ramifications not just
for quantum point contacts, but for all low-dimensional devices.

We now analyze the Reznikov {\em et al}. experiment
\cite{rez}.
A straightforward kinetic approach explains
both constant-voltage and constant-current results,
while automatically securing conservation.
We discuss why the nonequilibrium
noise of a quantum point contact cannot
be some kind of shot noise, as commonly believed:
it is a hot-electron fluctuation effect
unique to quasi-one-dimensional mesoscopic
conductors. Shot noise -- if present at all -- is
unimportant here.

Our treatment starts with orthodox quantum transport in metals
\cite{abri,gdii,dgjop}.
A uniform one-dimensional ballistic wire is intimately
contacted to large source and drain reservoirs.
The reservoir leads, permanently neutral and equilibrated, pin
the local chemical potential at each interface with the wire.
At the interfaces, the perturbed carrier distribution
goes smoothly to its equilibrium form in the leads,
set by the local chemical potential.

A generator forces an electron current into the wire
at the source and removes it through the drain.
In open-system conduction, the active injection and extraction
of external current is {\em necessary and sufficient}
to ensure charge conservation
not only microscopically, but globally
\cite{sols}.
This is a nontrivial and mandatory constraint
on any microscopic account of noise and transport. 

The external current evokes a resistivity-dipole field $E$ within
the uniform wire as the carrier density at both interfaces
adjusts to the disturbance.
The standard kinetic equation for the electron distribution
$f_k(t)$ is

\begin{widetext}
\begin{equation}
{{\partial f_k(t)}\over {\partial t}} +
{eE\over \hbar} {{\partial f_k(t)}\over {\partial k}}
= -{1\over {\tau_{\rm in}(\varepsilon_k)}}
{\left( f_k(t) -
{{\langle \tau_{\rm in}^{-1} f(t) \rangle}\over 
 {\langle \tau_{\rm in}^{-1} f^{\rm eq}(\mu, k_{\rm B}T) \rangle}}
f^{\rm eq}_k(\mu, k_{\rm B}T)
\right)}
- {1\over {\tau_{\rm el}(\varepsilon_k)}}
{ {f_k(t) - f_{-k}(t)}\over 2 }.
\label{eq1}
\end{equation}
\end{widetext}

\noindent
The inelastic and elastic relaxation rates, $1/\tau_{\rm in}$
and $1/\tau_{\rm el}$ respectively, parametrize the collision term.
They may depend on subband electron energy $\varepsilon_k$.
Expectations ${\langle ...\rangle}$ trace
over wavevector space
and spin; for example,
${\langle \tau_{\rm in}^{-1} f(t) \rangle}
= 2\int \tau_{\rm in}(\varepsilon_k)^{-1} f_k(t) dk/2\pi $.
The equilibrium function $f^{\rm eq}(\mu, k_{\rm B}T)$ has the usual
Fermi-Dirac form, dependent on chemical potential $\mu$
and thermal energy $k_{\rm B}T$.

Crucially, the collisional structures on the right of
Eq. (\ref{eq1}) secure charge and number conservation.
These essential properties are inherited
by the dynamical mean-square number fluctuation
$\Delta f_{\rm k}(t)$
in the structure, determining all the correlation
functions of physical interest
\cite{gdii,fnl}.
The equation for $\Delta f_{\rm k}(t)$
is systematically generated by varying
both sides of Eq. (\ref{eq1}).
Microscopic conservation is built in.
{\em So are the sum rules}
\cite{pinoz,thakur}.

In the low-frequency limit the noise spectral density in a QPC 
is duly obtained. For a device of operational length $L$,
subject to current $I$, it has the functional form

\begin{widetext}
\begin{equation}
S^{\rm xs}(I,\mu)
= \int^{\infty}_0 \!\!dt
\int^{L/2}_{-L/2} \!\!dx \int^{L/2}_{-L/2} \!\!dx'
{\Bigl( C_{JJ}(x-x'; I,\mu, t) - C_{JJ}(x-x'; I=0^+,\mu, t) \Bigr)};
\label{eq2}
\end{equation}
\end{widetext}

\noindent
we remove the equilibrium Johnson-Nyquist
noise floor ($I \to 0$).
The nonequilibrium current-current correlator
$C_{JJ}(x-x'; I,\mu, t)$
is computed from the Green function
for the spatially dependent version of Eq. (\ref{eq1}).
For further details of our QPC model see Ref.
\onlinecite{fnl}.

Equation (\ref{eq1}) and its variations are not
restricted to weak applied currents. $C_{JJ}$
is derived within a fully nonequilibrium description,
not only conserving but equally valid at strong
fields
\cite{gdii}
as at weak. The ability to do this,
in a strictly conserving way, is an absolute
prerequisite for analyzing the experiment of Ref.
\onlinecite{rez},
whose conditions take
a QPC far out of the weak-field, highly degenerate
limit addressed by Khlus
\cite{khlus}
and others
\cite{blbu}.

For collision times $\tau_{\rm el}$ and $\tau_{\rm in}$
independent of subband energy $\varepsilon_k$,
the nonequilibrium problem is exactly solvable
\cite{fnl}.
We now focus on a physical model for the behavior of
these times in a strongly nonequilibrium environment.

Let us start with elastic scattering.
Since the quantum point contact is impurity-free, its
elastic mean free path (MFP) is matched to $L$, the
operational length of the ballistic core.
The elastic scattering rate will not be
sensitive to the driving field.
The characteristic velocity of the carriers
is $v_{\rm av} \equiv {\langle |v| f^{\rm eq} \rangle}/n$
at the QPC electron density
$n = {\langle f^{\rm eq} \rangle}$.
Thus the elastic time is

\begin{equation}
\tau_{\rm el} = L/v_{\rm av}.
\label{eq3}
\end{equation}

\noindent
In the pinchoff limit, far below the subband threshold
$\varepsilon_1$, the density
$n \sim e^{-(\varepsilon_1 - \mu)/k_{\rm B}T}$ vanishes.
The carriers are classical; $v_{\rm av}$ is thermal.
For $\mu - \varepsilon_1 \gg k_{\rm B}T$
(degenerate limit) $v_{\rm av} \to v_{\rm F}$;
it is the Fermi velocity in the subband.

The behavior of $\tau_{\rm el}$ reflects the
constancy of the elastic MFP.
The inelastic MFP, however,
will decrease substantially with increasing 
source-drain voltage. The field-excited electrons
will shed excess energy by emitting many more phonons.
We model this inelastic loss via

\begin{equation}
\tau_{\rm in}(V, \mu)
= \tau_{\rm el}
{\left[
       {\varepsilon_{\rm av}\over \alpha eV}
\tanh{\left(
        {\alpha eV\over \varepsilon_{\rm av}}
     \right)}
\right]}^\beta.
\label{eq4}
\end{equation}   

\noindent
The driving voltage is $V = EL$ while
$\varepsilon_{\rm av} = {1\over 2}m^* v_{\rm av}^2$
is the characteristic energy of the subband population; it is
thermal near pinchoff and Fermi at high filling.
The ratio $eV/\varepsilon_{\rm av}$
determines the inelasticity. At high degeneracy,
$\varepsilon_{\rm av} = \mu - \varepsilon_1 \gg k_{B}T$
and Pauli exclusion inhibits phonon emission.
The parameters $\alpha$ and $\beta$ are set once
(to $\alpha = 0.52$ and $\beta = 0.6$)
to match the peaks at 250 and 300nA in Fig. \ref{f1}.
These values then determine all our results.

In the low-field limit, $\tau_{\rm in}(V, \mu) \to \tau_{\rm el}$.
This is the condition for ideal ballistic transport
in an open contact, when all dissipation is in the leads.
It yields Landauer's quantized
conductance steps across the subband thresholds
\cite{dgjop}.
As field-induced phonon emission sets in,
the inelastic MFP rapidly shrinks below $L$.
Eq. (\ref{eq4}) encodes this nonideal behavior.
 
\begin{figure}
  \centering
  \includegraphics[height=65mm]{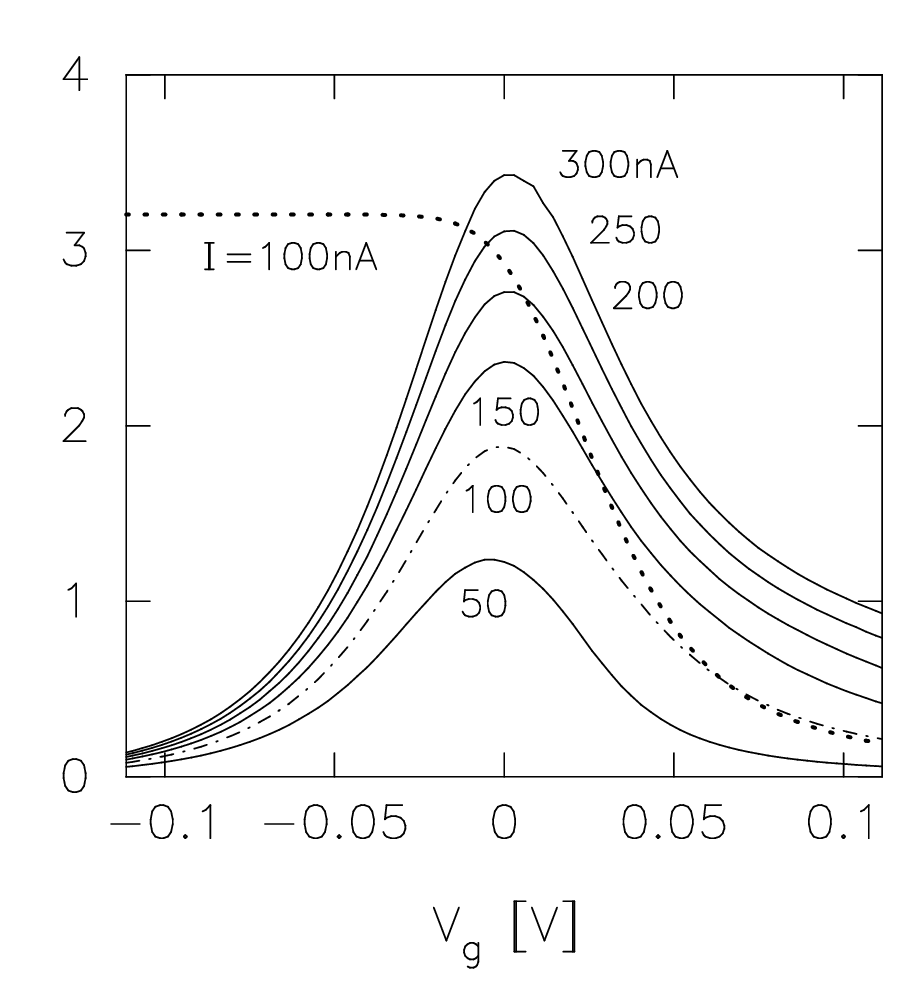}
  \caption{\label{f2}
Excess hot-electron noise at 1.5K in a QPC
at its first subband threshold, computed
with our strictly conserving Eq. (\ref{eq5}),
as a function of gate voltage
(relative to the first threshold), 
at fixed levels of source-drain current.
There is close quantitative affinity of
our peaks to the experimentally observed
first-threshold maxima in Fig. \ref{f1}
\cite{rez}.
Dotted line: corresponding prediction at $I = 100$nA
from a widely adopted theory of QPC noise
\cite{blbu},
using our transmissivity ${\cal T}_1$
(Eq. (\ref{eq6})) as
that model's phenomenologically required input.
Compare it with the kinetic result at 100nA (chain line).}
\end{figure}

In our parametrized inelastic model,
Eq. (\ref{eq2}) for the excess noise in the first
QPC subband becomes
\cite{fnl}

\begin{eqnarray}
S^{\rm xs}(I, \mu)
=&& \!\!\!
{{2e^2 I^2}\over m^*L^2}{1\over  G(I, \mu)}
\cr
&& \!\!\! \times
{\kappa\over \kappa_{\rm cl}}
{\left( \tau^2_{\rm in}
+ 2{\tau_{\rm el}\tau^2_{\rm in}\over {\tau_{\rm el}+\tau_{\rm in}}}
- {\tau^2_{\rm el}\tau^2_{\rm in}\over (\tau_{\rm el}+\tau_{\rm in})^2 }
\right)}.
\label{eq5}
\end{eqnarray}

\noindent
The conductance $G(I, \mu) = I/V$ is

\begin{equation}
G(I, \mu) \equiv {2e^2\over h}{\cal T}_1(I,\mu) =
{2e^2\over h}{\left( {v_{\rm F}\over v_{\rm av}} \right)}
{2\tau_{\rm in}\over {\tau_{\rm el}+\tau_{\rm in}}}.
\label{eq6}
\end{equation}

\noindent
At fixed current this determines the corresponding voltage.
The relation is nonlinear owing to the implicit
$V$-dependence through $\tau_{\rm in}$ from Eq. (\ref{eq4}).
Second, {\em as required by its sum rule}
\cite{thakur}, the compressibility
$\kappa = n^{-1}\partial \ln n/\partial \mu$
appears (in ratio with its classical value
$\kappa_{\rm cl} = 1/nk_{\rm B}T$).
The last factor on the right of Eq. (\ref{eq5})
is highly sensitive to the ratio
$\tau_{\rm in}/\tau_{\rm el}$ out of equilibrium
\cite{fnl}.

In Fig. \ref{f2} we display the result of
our conserving computation, implementing the physics
of Eqs. (\ref{eq3})--(\ref{eq6}).
Most dramatic are the peak structures
at the threshold $\mu = \varepsilon_1$
as the chemical potential takes $n$ in the channel
from low to high values.
The quantitative fit to the as-measured
peaks of Fig. \ref{eq1} is noteworthy.
The conventionally predicted shot-noise analog
\cite{blbu}
is entirely wide of the mark.

{\em What are the peaks}?
The peaks are a snapshot
of the carriers' transition from classical to quantum behavior.
They are generated by strongly competing trends within
$S^{\rm xs}(I, \mu)$: the compressibility $\kappa$, and
the combination of collision times in the numerator with
$G(I,\mu)$ in the denominator.
Consider the limiting cases.

(i) {\em Degenerate limit}.
Then $\kappa/\kappa_{\rm cl} \to k_{\rm B}T/2(\mu - \varepsilon_1) \ll 1$
even as the collision-time factor reaches
its ideal maximum of $1.75\tau_{\rm el}^2$
and $G$ plateaus out at $2e^2/h$. We have

\begin{equation}
S^{\rm xs}(I,\mu)
\propto (\kappa/\kappa_{\rm cl})(\tau_{\rm el}/L)^2 \sim n^{-4}. 
\label{eq7}
\end{equation}

(ii) {\em Pinchoff limit}.
At low densities, the carriers are classical.
The compressibility ratio $\kappa/\kappa_{\rm cl}$ goes to a
maximum of unity. At low $n$, though, carriers are highly
energized. Eq. (\ref{eq4}) tells us that

\begin{equation}
V = I/G \propto I/(n\tau_{\rm in}(V,\mu)) \to I/(nV^{-\beta}).
\label{eq7.5}
\end{equation}

\noindent
That is, $eV(I,\mu) \sim n^{-1/(1 - \beta)} \gg \varepsilon_{\rm av}$.
The direct dependence of $S^{\rm xs}$
on the collision times, along with $G \propto n\tau_{\rm in}$
in the denominator, then yields the overall asymptote

\begin{eqnarray}
S^{\rm xs}(I,\mu)
\propto&& (\kappa/\kappa_{\rm cl})\tau_{\rm in}(V,\mu)/n
{\left. \right.} \cr
\cr
\sim&& (n^{-1/(1 - \beta)})^{-\beta}/n
= n^{(2\beta - 1)/(1 - \beta)}.
\label{eq8}
\end{eqnarray}

\noindent
In either case Eqs. (\ref{eq7}) and (\ref{eq8}) lead to
vanishing shoulders. In the neighborhood of the threshold,
the peaks in $S^{\rm xs}(I,\mu)$ are the outcome
of the compressibility, heralding the onset of degeneracy,
playing against the dynamics of
nonequilibrium dissipative scattering in the channel.

\begin{figure}
  \centering
  \includegraphics[height=65mm]{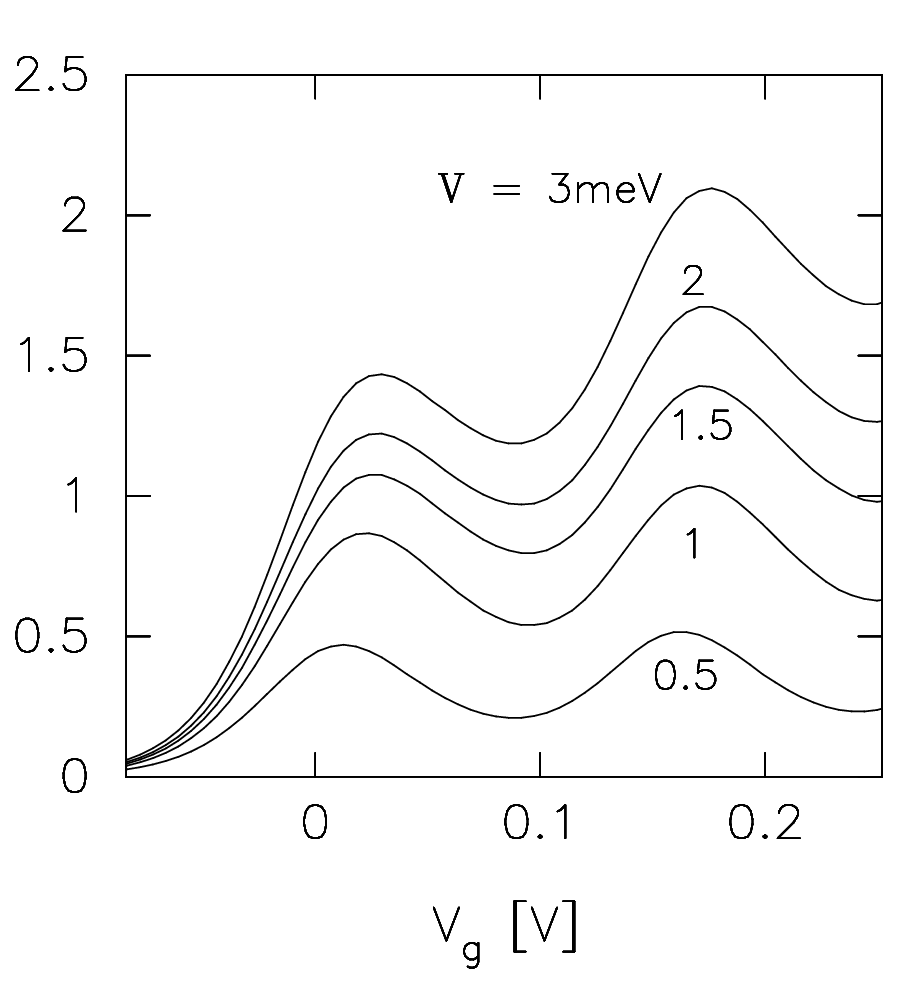}
  \caption{\label{f3}
Excess hot-electron noise in a QPC
at 1.5K for its first two subbands, at constant applied
voltage $V$, as a function of gate voltage
(referred to 1st threshold).
This corresponds to Fig. 2 of Ref. 
\onlinecite{rez}.
The quasi-linear dispersion of the peaks with $V$
shows that such dispersion is not unique to shot noise.}
\end{figure}

We present in Fig. \ref{f3}
the results of our noise model
at constant voltage.
Comparing it with the corresponding Fig. 2 of
Reznikov {\em et al.}
\cite{rez},
we again see a concordance between
observations and the kinetic calculation.

It is important to examine two aspects
of the Reznikov {\em et al.} noise data,
in light of what is needed to explain it.

(a) {\em Linear response is inapplicable}.
A quick estimate shows that the experiment
lies well beyond the limits of
weak-field linear models which perturb a
system mildly away from its Fermi surface
\cite{blbu}.
In the experimental regime of Fig. \ref{f3}
the gate voltage sweeps the channel's Fermi energy
from larger, safe values
(calibrating $G$ against measurements, we see that
$\mu \approx 3$meV covers three subbands),
far down to where carriers are not degenerate
at all, but classical.

In Fig. \ref{f3} the applied voltage goes up to 3meV. Since
$eV \gtrsim \mu$, linear response is clearly unjustified.
Much more extreme is Fig \ref{f2}, for constant current.
Eq. (\ref{eq7.5}) shows just how rapidly
$V$ reaches enormous values at channel densities
below threshold.

The severe limitation of weak-field models
has not inhibited their wide use in data analysis
\cite{rez,kumar,blbu}.
We stress that a properly constructed,
fully nonequilibrium kinetic approach -- one that
is strictly conserving -- is the only appropriate
analytical tool.

(b) {\em There is no shot noise}. Eq. (\ref{eq2})
and its specific form Eq. (\ref{eq5}) do not
describe shot noise. They describe hot-electron
noise, whose thermodynamic origin is different
\cite{gdii}.
This is borne out starkly by the peak structures
of Fig. \ref{f1}, in patent contradistinction to
shot-noise based predictions
\cite{blbu}.
Moreover, near-linearity of the peaks in Fig. \ref{f3}
with $V$ is not special to shot noise
\cite{rez}.
Such dispersion arises naturally
from inelastic dynamics.

Finally, readers will note that the noise plateaux
at very low densities in Fig. \ref{f1} are
not reproduced by those models that enforce
full shot noise (whose plateaux are far too big),
nor by the collisional model, in its presently simplified form.
What is really needed is a fuller theory that can
cross from the one-dimensionally confined QPC state
to one where the low-density, high-energy
electrons are so excited as to break the
confinement and fan out, nonuniformly
and at high momenta, into the drain.

We have presented a microscopically grounded
theory of nonequilibrium mesoscopic noise.
It is based on standard and rigidly conservative
kinetics. Thus it is well controlled,
and the whole range of relevant
noise properties of a quantum point contact
\cite{rez,kumar}
is well reproduced.
For the first time, we explain
qualitative and quantitative
experimental features that other theories
\cite{blbu}
miss entirely.

We have accounted for the prominent and enigmatic
noise peaks that have defied explanation until now.
The central role of the electronic compressibility
has been identified and quantified as
an essential physical determinant of
mesoscopic fluctuations. That is true not only
in QPCs but in all conductive structures at these scales.




\begin{thebibliography}{99}

\bibitem{rez}
M. Reznikov {\em et al}., Phys. Rev. Lett. {\bf 75}, 3340 (1995).

\bibitem{pinoz}
D. Pines and P. Nozi\`eres, {\em The Theory of Quantum Liquids},
(Benjamin, New York, 1966).

\bibitem{thakur}
J. S. Thakur {\em et al.}, Int. J. Mod. Phys. B {\bf 18}, 1479 (2004);
see also arXiv preprint cond-mat/0401134.

\bibitem{kumar}
A. Kumar {\em et al}., Phys. Rev. Lett. {\bf 76}, 2778 (1996).

\bibitem{khlus}
V. A. Khlus, Zh. \'Eksp. Teor. Fiz. {\bf 93}, 2179 (1987)
[Sov. Phys. JETP {\bf 66}, 1243 (1987)].

\bibitem{blbu}
Y. M. Blanter and M. B\"uttiker, Phys. Rep. {\bf 336}, 1 (2000).

\bibitem{abri}
A. A. Abrikosov, {\em Fundamentals of the Theory of Metals}
(North-Holland, Amsterdam, 1988).

\bibitem{gdii}
F. Green and M. P. Das, J. Phys.: Condens. Matter
{\bf 12}, 5251 (2000).

\bibitem{dgjop}
M. P. Das and F. Green, J. Phys.: Condens. Matter
{\bf 15}, L687 (2003).

\bibitem{sols}
F. Sols, Phys. Rev. Lett. {\bf 67}, 2874 (1991).

\bibitem{fnl}
F. Green and M. P. Das, Fluct. Noise Lett. {\bf 1}, 21 (2001).


\end{thebibliography}
\end{document}